\documentclass{eptcs}
\providecommand{\event}{DCM 2009}
\providecommand{\volume}{0}
\providecommand{\anno}{2009}
\providecommand{\firstpage}{1}
  \usepackage[latin1]{inputenc}
  \usepackage{graphicx}
  \usepackage{color}
  \usepackage{amsfonts}
  \usepackage{amssymb}
  \usepackage{amsmath}
  \usepackage{verbatim}
  \usepackage{epic, eepic}
  \usepackage{breakurl}
  \newtheorem{theorem}{Theorem}
  \newtheorem{definition}{Definition}
  \newtheorem{proposition}{Proposition}
  
  \newtheorem{example}{Example}
  \newcommand{\virg}[1]{``#1''}
  \newcommand{\preco}[1]{\mbox{${}^\bullet{#1}$}} 
  \newcommand{\postc}[1]{\mbox{${#1}^\bullet$}} 
  \newcommand{\ort}{\perp}                       
  \newcommand{\pang}[1]{\langle#1\rangle}        
  \newcommand{\aarg}{\,.\,}                     
  \title{Orthomodular Lattices Induced by the Concurrency Relation}
  \author{Luca Bernardinello \and
          Lucia Pomello
  \institute{Dipartimento di informatica, sistemistica e comunicazione\\
             Universit\`a degli studi di Milano--Bicocca\\
             viale Sarca 336--U14, Milano, Italy}
             \email{luca.bernardinello@unimib.it} \and
          Stefania Rombol\`a}
\def\titlerunning{Orthomodular Lattices}
\def\authorrunning{Luca Bernardinello \& Lucia Pomello \& Stefania Rombol\`a}
\begin{document}
 \maketitle
\begin{abstract}
We apply to locally finite
partially ordered sets a construction which associates
a complete lattice to a given poset;
the elements of the lattice are 
the closed subsets of a closure
operator, defined starting from the \emph{concurrency} relation.
We  show that, if the partially ordered set satisfies 
a property of local density, i.e.: N-density, 
then the associated lattice is also orthomodular.
We then  consider occurrence
nets, introduced by C.A. Petri as models of concurrent computations, and define
a family of subsets of the elements of an occurrence net;
we call those subsets \emph{causally closed} because they
can be seen as subprocesses of the whole net which are,
intuitively, closed with respect to the forward and backward local state changes.
We show that, when the net is K-dense, the causally
closed sets coincide with the closed sets induced by the
closure operator defined starting from the
concurrency relation. K-density is a property of partially ordered sets introduced by Petri,
on the basis of former axiomatizations of special relativity theory.

\end{abstract}
%
%
  \section{Introduction}
%
We consider models of concurrent behaviours based on partial orders, and in particular 
occurrence nets. 
Occurrence nets were introduced by C.A. Petri (\cite{P77}) as a model of
non sequential processes which are physically realizable.
They are a special kind of Petri nets,
where occurrences of local states (also called conditions) and of events are partially
ordered.
The partial order is interpreted as a kind of causal dependence relation
(for background on partial orders and occurrence nets, see~\cite{BF88}).

Any partially ordered set (or \emph{poset} for short) induces a \emph{concurrency} relation, defined as the complement of
the partial order. This relation is symmetric but in general non transitive.

By applying known techniques of lattice theory, one can derive, from such a relation,
a complete, orthocomplemented lattice of subsets of the underlying set.

In a recent paper (\cite{BPR09}) we showed that this technique, applied to an
occurrence net, always gives an orthomodular lattice.

In the present paper, we consider N-density and K-density, properties defined by Petri,
inspired by former axiomatizations of special relativity theory
(see, for example, \cite{C58}).
A partial order is K-dense if any \emph{line} (namely, a maximal subset of pairwise
ordered elements) intersects any \emph{cut} (a maximal subset of pairwise
incomparable elements).
This corresponds to the intuitive idea that in a given global state (represented by a
cut) any sequential subprocess is in some state, given by a point along the
subprocess.
N-density is a sort of local, and weaker, form of K-density.

We show that N-density, together with two local finiteness properties, of a poset is
sufficient to produce an orthomodular lattice.
This generalizes one of the main results of~\cite{BPR09}.

We then restrict attention to degree-finite and interval-finite occurrence nets.
On these nets we introduce the notion of \emph{causally closed set}, 
which corresponds to a set of elements of the net 
which identifies a sort of  \emph{causally closed} subprocess, 
i.e.: a subprocess uniquely constructible starting from a set of concurrent conditions.
We show that closed sets, as defined in~\cite{BPR09},  are causally closed and prove that, 
in the case of K-dense occurrence nets, closed sets and causally closed sets coincide.

The next section collects some definitions and results to be used
later. In Section~\ref{s:chiusico}, we show that N-density and
interval-finiteness suffice to grant the orthomodularity
of the lattice of closed sets generated starting from the
concurrency relation. Section~\ref{s:subproc} introduces the notion
of causally closed set, and shows that in K-dense occurrence nets,
closed sets and causally closed sets coincide.

Proofs are omitted, but can be found in~\cite{bpr09dcmex}.
\section{Preliminary Definitions}\label{s:preldef}

\subsection{Orthomodular Posets and Lattices}
In this section we recall the basic definitions
related to orthomodular posets and lattices.
\begin{definition}      
  An \emph{orthocomplemented poset}
  ${\cal P}=\pang{P, \leq, 0, 1,(\aarg)'}$
  is a partially ordered set $(P, \leq)$, equipped with a
  minimum and a maximum element, respectively denoted by 0 and
  1, and with a map $(\aarg)':P \rightarrow P$,
  such that the following conditions are verified (where
  $\lor$ and $\land$ denote, respectively, the least
  upper bound and the greatest lower bound with respect to
  $ \leq $, when they exist):
  $ \forall x, y \in P $
\begin{align*}
    &\emph{(i)} \quad (x')'      = x;\\
      &\emph{(ii)} \quad x \leq y  \Rightarrow y' \leq x';\\
      &\emph{(iii)} \quad x \land x'  = 0  \ \emph{and} \ x \lor x'  = 1.
\end{align*}
\end{definition}
The map $(\aarg)':P \rightarrow P$ is 
called an \emph{orthocomplementation} in $P$.
In an orthocomplemented poset, $\land$ and $\lor$, when they exist,
are not independent: 
in fact, the so-called De Morgan laws hold:
$(x \lor y)'= x' \land y'$, $(x \land y)'= x' \lor y'.$
In the following, we will sometimes use \emph{meet} and
\emph{join} to denote, respectively, $\land$ and $\lor$.
Meet and join can be extended to families of elements in the obvious way,
denoted by $\bigwedge$ and $\bigvee$.
\begin{figure}
\begin{center}
\includegraphics[width=6.5cm]{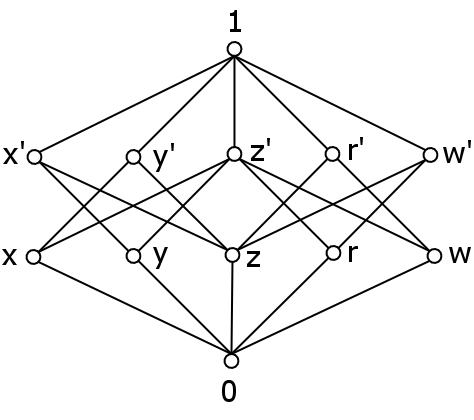}
\caption{A finite orthomodular lattice.}\label{f:ortomodulare}
\end{center}
\end{figure}

A poset $(P,\leq)$ is called \emph{orthocomplete} when it is orthocomplemented and
every countable subset of pairwise orthogonal elements of
$P$ has a least upper bound.

A lattice $\mathcal{L}$ is a poset 
in which for any pair of elements meet and join always exist. 
A lattice $\mathcal{L}$ is \emph{complete} when the meet and the join
of any subset of $\mathcal{L}$ always exist. 

\begin{definition}\cite{BC81}
  An \emph{orthomodular poset} ${\cal P}=\pang{P, \leq, 0, 1,(\aarg)'}$
  is an orthocomplete poset which satisfies the condition:
  \begin{displaymath}
    x \leq y  \Rightarrow y = x \lor (y \land x')
  \end{displaymath}
  \end{definition}
  which is usually referred to as the \emph{orthomodular} law.
  
The orthomodular law is weaker than the distributive law.
A lattice $\mathcal{L}$ is called \emph{distributive} if and only if $\forall x,y,z \in \mathcal{L}$
the equalities
$x \land (y \lor z) = (x \land y) \lor (x \land z)$,
$x \lor(y \land z) = (x \lor y) \land (x \lor z)$
hold.
Orthocomplemented distributive lattices are called Boolean algebras.
Orthomodular posets and lattices can therefore be considered as a generalization of Boolean
algebras and have been studied as algebraic models for quantum logic \cite{PP91}.

Any orthomodular lattice can be seen as a family of
partially overlapping Boolean algebras.
Figure~\ref{f:ortomodulare} shows a finite orthomodular lattice.


\subsection{Closure Operators}\label{s:closop}

\begin{definition}
Let $X$ be a set and $\mathbb{P}(X)$ the powerset of $X$.
A map $\mathcal{C}: \mathbb{P}(X) \rightarrow \mathbb{P}(X)$ 
is a \emph{closure operator} on $X$ if, for all $A,B \subseteq X$,
\begin{align*}
&\emph{(i)} \quad A \subseteq \mathcal{C}(A),\\
&\emph{(ii)} \quad A  \subseteq B \Rightarrow \mathcal{C}(A) \subseteq \mathcal{C}(B),\\
&\emph{(iii)} \quad \mathcal{C}(\mathcal{C}(A))=\mathcal{C}(A).
\end{align*}
\end{definition}
A subset $A$ of $X$ is called \emph{closed} with respect to $\mathcal{C}$ 
if $\mathcal{C}(A)=A$.
If $\mathcal{C}$ is a closure operator on a set $X$, 
the family
$\{A \subseteq X \ | \ \mathcal{C}(A)=A\}$
of closed subsets of $X$ forms a complete lattice, 
when ordered by inclusion, in which
\begin{displaymath}
\bigwedge \{A_i: i \in I\}= \bigcap_{i \in I} A_i,  \quad
\bigvee \{A_i: i \in I\}= \mathcal{C}(\bigcup_{i \in I} A_i).
\end{displaymath}
The proof of this statement can be found in \cite{B79}.

We now describe a well-known construction from binary relations
to closure operators (see, for example, \cite{B79}).
Let $X$ be a set, 
and $\alpha \subseteq X \times X$ be a symmetric relation. 
Define an operator $(.)^\ort$ on the powerset of $X$:
given $A \subseteq X$ 
\begin{displaymath} 
A^\ort=\{x \in X \ | \ \forall y \in A: (x,y) \in \alpha\}.
\end{displaymath}

By applying twice the operator $(\aarg)^\ort$,
we get a new operator $C(\aarg)=(\aarg)^{\ort \ort}$.
The map $C$ on the powerset of $X$ is a closure operator on $X$.
A subset $A$ of $X$ is called \emph{closed} 
with respect to $(\aarg)^{\ort \ort}$ 
if $A=A^{\ort \ort}$. 
The family $L(X)$
of all closed sets of $X$, 
ordered by set inclusion, is a complete lattice.

When $\alpha$ is also irreflexive, 
the operator $(\aarg)^\ort$ applied to elements of $L(X)$ 
is an orthocomplementation; the structure 
$\mathcal{L}(X)=\pang{L(X), \subseteq, \emptyset, X, (\aarg)^\ort}$ 
then forms an orthocomplemented complete lattice.

\subsection{Occurrence Nets}

\begin{definition} 
  A \emph{net} is a triple $ N = (B, E, F) $, where
    $B$ and $E$ are countable sets,
    $ F \subseteq ( B \times E ) \cup ( E \times B) $, and
    \begin{itemize}
      \item [\emph{(i)}] $ B \cap E = \emptyset $
       \item [\emph{(ii)}]  $ \forall e \in E \ \  \exists x, y \in B : (x,e) \in F $ and $(e,y) \in F$.
    \end{itemize}
 \end{definition}   
The elements of $B$ are called \emph{local states} or \emph{conditions}, 
the elements of $E$ \emph{local changes of state} or \emph{events}, 
and $F$ is called the \emph{flow relation}.  
Note that we allow isolated conditions but not isolated events.

Local states correspond to properties which can be true or false
in a given global state of the system;
potential global states of the system
modeled by $N$ are subsets of local states.

For each $x \in B \cup E$, define
  $ \preco{x} = \{ y \in B \cup E \ | \ (y, x) \in F \} $,
  $ \postc{x} = \{ y \in B \cup E \ | \ (x, y) \in F \} $.
For $e \in E$, an element $b \in B$ is a \emph{precondition} of 
$e$ if $b \in \preco e$; it is a \emph{postcondition} of $e$
if $b \in \postc e$.

Occurrences of events are in accord with the following \emph{firing rule}:
an event may occur if its preconditions are true and
its postconditions are false;
when the event occurs its preconditions become false, while its postconditions become true.
In this way the occurrence of an event changes the current global state 
by only changing the \emph{local} states directly connected to the event itself.


Occurrence nets are a special class of nets 
used to model non-sequential processes (\cite{P77}, \cite{BF88}) by recording the partial order of the validity of conditions and of the  event occurrences in the evolution of the system behaviour.

\begin{definition}\label{reteoccorrenze}
A net $N=(B,E,F)$ is an \emph{occurrence net} iff
\begin{itemize}
\item [\emph{(i)}] $\forall b \in B: |\preco b| \le 1 \ \land \ |\postc b| \le 1$ and
\item [\emph{(ii)}]$\forall x,y \in B \cup E:(x,y) \in F^+ \Rightarrow (y,x) \notin F^+$,
\end{itemize}
where $F^+$ denotes the transitive closure of $F$.
\end{definition}
Definition \ref{reteoccorrenze}(i) means that 
an occurrence net does not contain non-deterministic choices, 
the idea being that all conflicts are resolved at the behavioural level.
Definition \ref{reteoccorrenze}(ii) means that 
an occurrence net contains no cycles, 
the idea being that all loops are unfolded at the behavioural level.

Because of Definition \ref{reteoccorrenze}(ii), 
the structure $(X,\sqsubseteq)$ derived from an occurrence net $N$ 
by putting $X=B \cup E$ and $\sqsubseteq=F^*$
($F^*$ denotes the reflexive and transitive closure of $F$)
is a partially ordered set (shortly \emph{poset}).
We will use $\sqsubset$ to denote the associated 
strict partial order. 

Given a partial order relation $\leq$ on a set P,
we can derive the relations $li= \ \leq \cup \geq$,
and $co=(P \times P)\setminus li$.
We will be interested in such relations derived from
$(X, \sqsubseteq)$. In such case,
intuitively, $x \ li \ y$ means that $x$ and $y$ 
are connected by a causal relation,
and $x \ co \ y$ means that $x$ and $y$ 
are causally independent. 
The relations $li$ and $co$ are symmetric and not transitive. 
Note that
$li$ is a reflexive relation, while $co$ is irreflexive.
Given an element $x \in X$ and a set $S \subseteq X$, 
we write $ \ x \ co \ S$ 
if $\forall y \in S: x \ co \ y$.
Moreover, given two sets $S_1 \subseteq X$ and $S_2 \subseteq X$, 
we write $S_1 \ co \ S_2$ if $\forall x \in S_1, \forall y \in S_2: x \ co \ y$.
In the following we will use $x\, co\, y$ or $(x,y) \in co$
indifferently, and similarly for $li$.

On the basis of the flow relation $F$ and its transitive closure $F^+$,
to each $x \in X$ we associate the elements in its \emph{past}  and in its \emph{future}, denoted by:
\begin{align*}
&F^-(x) = \{ y \in X \ | \ y \sqsubset x \} \ \textrm{and}\\
&F^+(x) = \{ z \in X \ | \ x \sqsubset z \} 
\end{align*}
 respectively. 
By generalizing to subsets $S$ of $X$, 
we denote the \emph{past} and the \emph{future} of $S$ by 
\begin{align*}
&F^-(S)=\{x \in X \ | 
\ x \notin S, \exists y \in S: x \in F^-(y)\} \ \textrm{and} \\
&F^+(S)=\{x \in X \ | 
\ x \notin S, \exists y \in S: x \in F^+(y)\}.
\end{align*}
From the definition, it follows that an element $x$
belongs neither to its future nor to its past. 

A \emph{clique} of a binary relation is a set of pairwise related elements.
From the $co$ and $li$  relations one can define \emph{cuts} and
\emph{lines} of a poset $\mathcal{P} = (P, \leq)$ as 
maximal cliques of $co$ and $li$, respectively:

\begin{align*}
&\mathrm{Cuts}(\mathcal{P})=
     \{c \subseteq P \ | \ c \ \textrm{is a maximal clique of } co \cup id_P \};\\
&\mathrm{Lines}(\mathcal{P})=
     \{\ l \subseteq P \ | \ l \ \textrm{is a maximal clique of } li\}.
\end{align*}
Given an occurrence net $N = (B, E, F)$, we will denote by
$\mathrm{Cuts}(N)$ and $\mathrm{Lines}(N)$, respectively,
the set of cuts and the set of lines of the poset associated
to $N$. 
We will always assume the Axiom of Choice, so that any clique
of $co$ and of $li$ can be extended to a maximal clique.
In particular, we will denote cliques and maximal cliques of $co$ 
which  contain only conditions by $B$-cosets and $B$-cuts, respectively.

C.A. Petri formalized some properties
which intuitively should hold for posets corresponding to
non-sequential processes which are actually feasible \cite{P80}, see also \cite{BF88}.
In particular, we will consider interval-finiteness,
degree-finiteness, a sort of local density called N-density,
and K-density.

\begin{definition}
$\mathcal{P} = (P, \leq)$ is \emph{interval-finite}
$\Leftrightarrow \forall x,y \in P: |[x,y]|< \infty$, 
where  $[x,y]=\{z \in P \ | \ x \leq z \leq y\}$.
\end{definition}
For $x, y \in P$, we write $x \lessdot y$ if $x < y$ and,
for all $z \in P$, $x < z \leq y \Rightarrow z = y$.
Let $\preco{x} = \{\, y\,|\, y \lessdot x\,\}$,
and $\postc{x} = \{\, y\,|\, x \lessdot y\,\}$.
\begin{definition}
$\mathcal{P} = (P, \leq)$ is \emph{degree-finite}
$\Leftrightarrow \forall x \in P:  
|\preco x| < \infty \ \mathrm{and}\ |\postc x| < \infty$.
\end{definition}

\begin{definition}
$\mathcal{P} = (P, \leq)$ is \emph{N-dense}
$\Leftrightarrow \forall \ x,y,v,w \in P$:
$(y < v$ and $y < x$ and $w < v$ and
$(y \ co \ w \ co \ x \ co \ v)) \Rightarrow \exists z \in P: (y < z < v$
and
$(w \ co \ z \ co \ x))$.
\end{definition}

\begin{figure}
\begin{center}
\includegraphics[width=8cm]{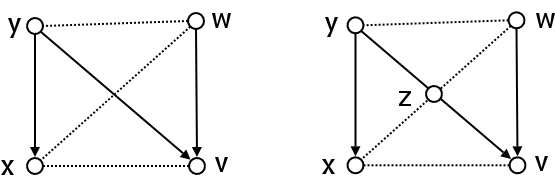}
\caption{Illustration of N-density.}\label{f:Ndense}
\end{center}
\end{figure}

For a graphical representation of N-density condition see Figure \ref{f:Ndense}.

\begin{proposition}\label{p:Netocc_Ndense} \cite{BF88}
Let $(X,\sqsubseteq)$ be the poset associated to an occurrence net $N=(B,E,F),
\ X=(B \cup E)$.
Then $(X,\sqsubseteq)$ is N-dense.
\end{proposition}

K-density is based on the idea of interpreting 
cuts as (global) states and lines as sequential subprocesses.
K-density postulates that every occurrence of a subprocess 
must be in some state.

\begin{definition}
$\mathcal{P} = (P, \leq) \ \textrm{is} \ \emph{K-dense} \Leftrightarrow 
\forall c \in \mathrm{Cuts}(\mathcal{P}),
\forall l \in \mathrm{Lines}(\mathcal{P}): 
c \cap l \neq \emptyset$. 
\end{definition}
Obviously, in general $|c \cap l| \leq 1$.
An occurrence net $N$ is K-dense if its associated poset
is K-dense.

\section{Closed Sets Induced by the Concurrency Relation}\label{s:chiusico}
In this section we apply the construction recalled at the
end of Section~\ref{s:closop}
to the $co$ relation in partially ordered sets,
and study the resulting algebraic structure of closed sets.

Let $\mathcal{P} = (P, \le)$ be a poset. We can define
an operator on subsets of $P$, 
which corresponds to an orthocomplementation,
since $co$ is irreflexive, and by this operator we define closed sets.

\begin{definition}\label{d:coclosed}
Let $S \subseteq P$, then
\begin{align*}
&(i) \quad S^\ort=\{x \in P \ | \ \forall y \in S: x \ co \ y\} \
 \textrm{is the \emph{orthocomplement} of} \ S; \\
 &(ii) \quad \textrm{if } S = (S^\perp)^\perp,
       \textrm{then $S$ is a \emph{closed set} of} \ P.
\end{align*} 
\end{definition}

The set $S^\ort$ 
contains the elements of $P$ 
which are not in causal relation with any element of $S$.
Obviously, $S \cap S^{\ort} = \emptyset$ for any $S \subseteq P$.
In the following, we sometimes denote 
$(S^\perp)^\perp$ by $S^{\perp \perp}$.
Note that:
$\forall c \in Cuts(\mathcal{P})$,  $c^\ort= \emptyset$ and $ c^{\ort\ort}=P$.

\begin{figure}
\begin{center}
\includegraphics[width=6cm]{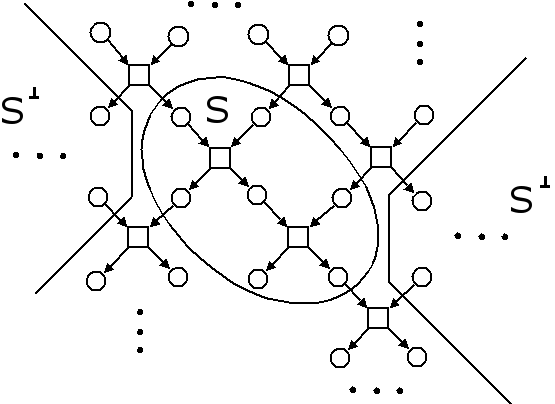}
\caption{A closed set $S$ and its orthocomplement $S^\ort $.}\label{f:chiuso}
\end{center}
\end{figure}

\begin{example}
  Figure \ref{f:chiuso} shows a closed set $S$ and
  its orthocomplement $S^\ort $ on the poset associated
  to an occurrence net. 
\end{example}
A closed set $S$ and its orthocomplement $S^\ort$ share the past
and the future.

\begin{proposition} \cite{BPR09}\label{p:ortocomplementi} 
Let $S=S^{\ort \ort}$. Then $F^-(S) = F^-(S^\ort)$ and $F^+(S) = F^+(S^\ort)$. 
\end{proposition}


Now we study the algebraic structure induced by 
the closure operator defined above.
We call $L(P)$ the collection of closed sets of $\mathcal{P}=(P,\le)$. 
By the results on closure operators recalled in Section~\ref{s:closop},
we know that
\[
  \mathcal{L}(P)= \pang{L(P), \subseteq, \emptyset, P, (\aarg)^\ort}
\]
is an orthocomplemented complete lattice, in which the meet is just
set intersection, while the join of a family of elements
is given by set union followed by closure. 

In general, the structure $\mathcal{L}(P)$ is not orthomodular and 
a fortiori non distributive, as can be seen in the following example.

\begin{example}
Let us consider the poset $\mathcal{P}=(P, \le)$ shown
in the left side of Fig.~\ref{f:Ndense}
and the closed sets $\{w\}$ and $ \{v,w\}$;
$\{w\}^\ort=\{x,y\}$.
The orthomodular law is not valid since
$\{w\}\subset\{v,w\}, (\{x,y\} \land \{v,w\})=\emptyset$
and hence $(\{w\} \lor \emptyset)$
 is not equal to $\{v,w\}$.
\end{example}

The poset considered in the previous example is not N-dense.
It is natural to investigate if there is a relation between
N-density of a poset $P$ and the orthomodularity of the associated 
structure of closed sets $\mathcal{L}(P)$.
It turns out that N-density of an interval-finite poset
is a sufficient condition even if not necessary for orthomodularity.
 
\begin{theorem}\label{t:Ndense_Orth}
Let $\mathcal{P}=(P,\le)$ be an N-dense, interval-finite poset. 
Then $\mathcal{L}(P)$ is an orthomodular lattice.  
\end{theorem}

\begin{figure}
\begin{center}
\includegraphics[width=3.1cm]{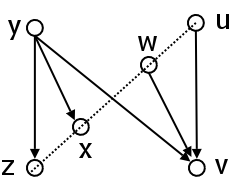}
\caption{\label{f:noNdense_Orth}A non N-dense poset generating an orthomodular lattice.}
\end{center}
\end{figure}

The reverse implication is not true, as can be seen in the following example.
\begin{example}
Figure \ref{f:noNdense_Orth} shows a poset $\mathcal{P}=(P,\le)$ which is not N-dense.
The family of closed sets $L(P)$ forms a Boolean,
hence orthomodular, lattice, whose atoms are
$\{z\}, \{x\},\{w\},\{u\}$.
\end{example}

From Theorem \ref{t:Ndense_Orth} and Proposition \ref{p:Netocc_Ndense} it follows 
that the family $L(N)$ of closed sets of the poset $(X, \sqsubseteq)$ associated to 
an interval-finite occurrence net $N=(B,E,F)$, with $X=(B \cup E)$,
forms an orthomodular lattice
$\mathcal{L}(N)=\pang{L(N), \subseteq,$
$\emptyset, X, (\aarg)\ort}$.
This result  constitutes a generalization and a different proof of a theorem  in \cite{BPR09}, 
where interval-finite and degree-finite occurrence nets have
been considered and where closed sets have been called ``causally closed sets''.

\section{Causally Closed Sets in Occurrence Nets}
\label{s:subproc}
In this section we consider only interval-finite and degree-finite occurrence nets.
We introduce particular subsets of elements of such an occurrence net $N=(B,E,F)$,
which we call \virg{causally closed sets}, since they can be interpreted as subprocesses 
which can be uniquely obtained by a particular $B$-coset. 

We show that,  for K-dense occurrence nets,  \virg{causally closed sets}  coincide with
closed sets of the poset associated to the occurrence net, as defined in the previous section.

\begin{definition} \label{d:sottoinsiemi_causalmente_chiusi}
Let $N=(B,E,F)$ be an occurrence net. 
$C \subseteq B \cup E$ is a \emph{causally closed set} iff
\begin{itemize}
\item [\emph{(i)}] $\forall e \in E$, $\preco e \subseteq C \Rightarrow
e \in C$,
\item [\emph{(ii)}] $\forall e \in E$, $\postc e \subseteq C \Rightarrow
e \in C$,
\item [\emph{(iii)}] $\forall e \in E$, $e \in C \Rightarrow
\preco e  \cup \postc e \subseteq C$,
\item [\emph{(iv)}] $\forall x,y \in C$, $x \ li \ y \Rightarrow
[x,y] \subseteq C$.
\end{itemize}
\end{definition}

A causally closed set is therefore a convex set which,  intuitively, is  closed with respect to the firing rule, 
i.e.: if it contains an event then it contains also all its preconditions and postconditions, 
and, moreover, if it contains all the preconditions or all the postconditions of an event 
then it contains the event itself.

\begin{figure}
\begin{center}
  \hspace*{\fill}
  \includegraphics[width=5.2 cm]{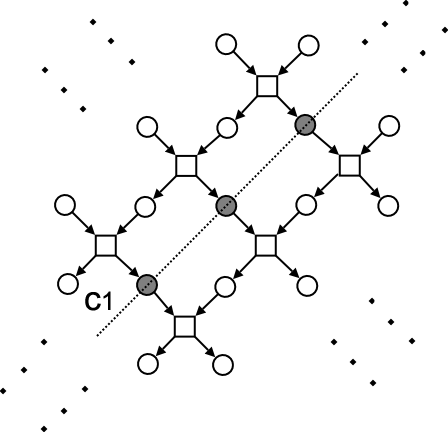} \hfill
  \includegraphics[width=5.6 cm]{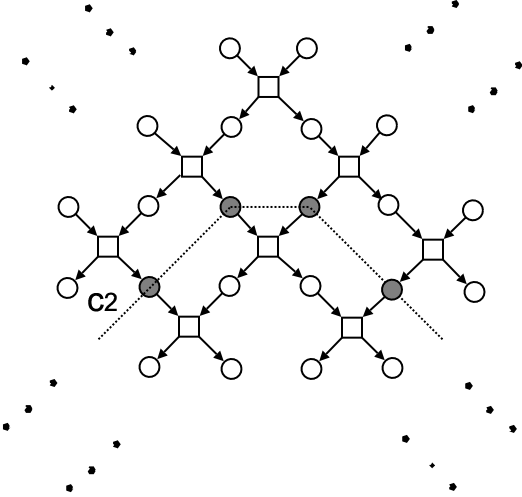}\hspace*{\fill}
  \caption{\label{f:griglia}Examples illustrating that in general $\phi(A) \neq A^{\ort \ort}$.}
\end{center}
\end{figure}

\begin{example}
The set $S$ in Figure~\ref{f:chiuso} is causally closed.
The set of grayed elements on the left side of Figure~\ref{f:griglia}
  is causally closed.
  On the contrary, the set of grayed elements on the right side of Figure~\ref{f:griglia} 
  is not causally closed because, for example, it contains the preconditions of an event 
  but it does not contain the event itself.
  \end{example}
We call $CC(N)$ the family of causally closed sets of $N$.
It is easy to prove that $CC(N)$ is closed by intersection,
$\emptyset \in CC(N)$, $B \cup E \in CC(N)$.
Hence the family $CC(N)$  forms a complete lattice,
where meet is given by set intersection.
In general this lattice is not orthocomplemented.

\begin{definition}\label{d:costruzione_sottoinsiemi_causalmente_chiusi}
Let $N=(B,E,F)$ be an occurrence net,  $X = B \cup E$,  and  $\mathbb{P}(X)$ be the powerset of $X$.
Define $\phi: \mathbb{P}(X) \rightarrow \mathbb{P}(X)$ as follows:
$\forall A \subseteq X$,
$\phi(A)=\bigcap \{C_i \ | \ C_i \in CC(N)$ and $A \subseteq C_i\}$.
\end{definition}
\noindent
Note that  $\phi$ is a closure operator, 
$A \subseteq \phi(A)$ and the frontier of
$\phi(A)$ is a subset of conditions, 
where  the frontier of a subset $A$ of $X$ is the set 
$\{ x \in A  \ | \  \exists y \in X \setminus A$
such that:  $x F y $ or  $y Fx \}$.

Closed sets, as defined in Section \ref{s:chiusico},
are causally closed sets, and this directly follows
from a characterization of closed sets given in \cite{BPR09}.
However, in general a causally closed set is not a closed set,
in fact the following example shows two cases in which $\phi(A) \neq A^{\ort \ort}$.
\begin{example} Figure \ref{f:griglia} shows a non K-dense occurrence net. 
Remember that
$\forall c \in Cuts(N), c^\ort= \emptyset$ and $ c^{\ort\ort}=X$.
Consider now the B-cut $c_1$ formed by the gray conditions
on the left side of the figure;
$\phi(c_1)=c_1$.
For the B-cut $c_2$ on the right side of the figure,
$c_2 \subset \phi(c_2) \subset X$.
\end{example}
When $A$ is a B-coset of $N=(B,E,F)$,
$\phi(A)$ can be inductively constructed,
as shown in the following.
\begin{definition} \label{d:costruzione_Ai}
Given $A_i \subseteq B \cup E$,
define $A_{i+1}= A_i \cup \{Int(e) \ | 
\ e \in E, \preco e \subseteq A_i \lor \postc e \subseteq A_i\}$,
where for $e \in E, Int(e)=\{e\} \cup \preco e \cup \postc e$.
\end{definition}
Note that $A_i \subseteq A_{i+1}$ for every $i$.

\begin{proposition} \label{p:costruzione_chiuso}
Let $A$ be a $B$-coset of $N$.
Then $\bigcup_{i \in \mathbb{N}} A_i = \phi(A)$, where $A_0=A$.
\end{proposition}

\begin{figure}
  \hspace*{\fill}
  \includegraphics[width=0.47\linewidth]{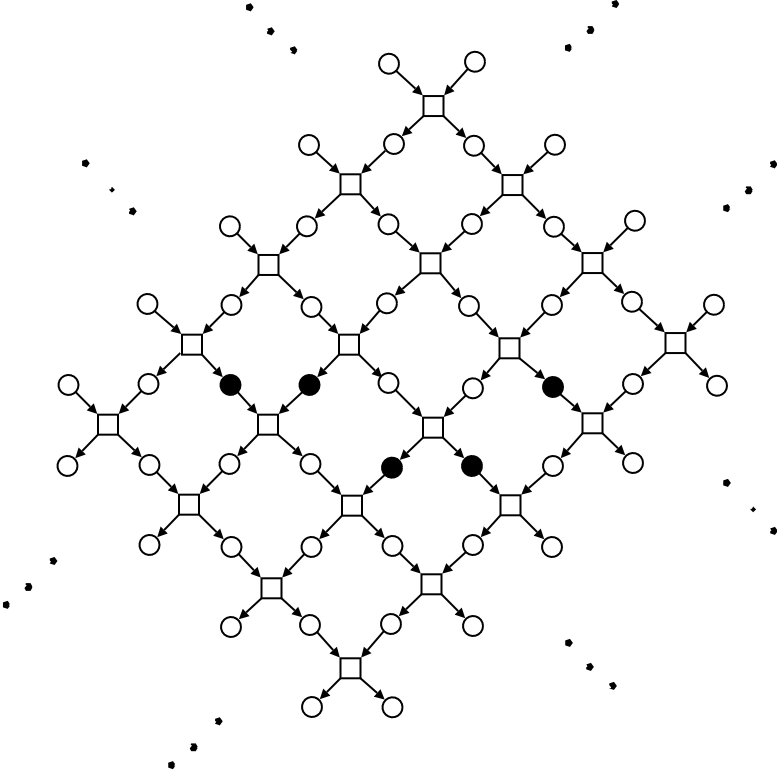}\hfill
  \includegraphics[width=0.47\linewidth]{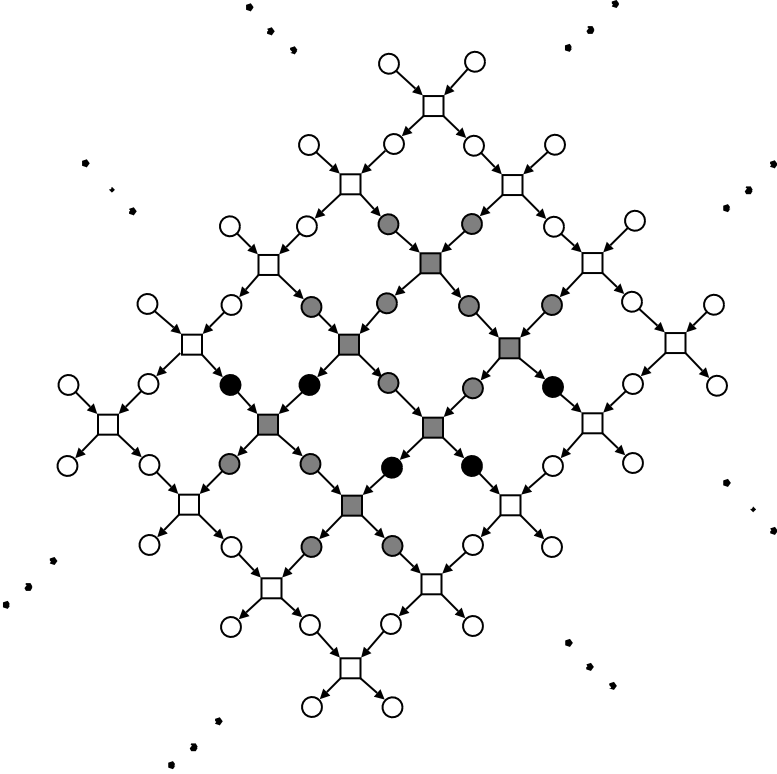}\hspace*{\fill}
  \caption{Inductive construction of a causally closed set.}
  \label{f:ind_const}
\end{figure}

The inductive construction of a causally closed set from
a B-coset is shown in Figure~\ref{f:ind_const}.
This motivates the name of 
$\emph{causally closed}$ sets and suggests  an interpretation of these as 
$\emph{non sequential}$ $\emph{subprocesses}$, which are $\emph{causally closed}$ 
in the sense that are uniquely constructed starting by a B-coset of the whole process.
The construction, in fact, simulates the forward and backward run of the system  starting from a B-coset:
it adds  all the events such that either all their preconditions 
or all their postconditions belong to the starting B-coset,
and then it proceeds by adding to the set all the post/preconditions of the added events,
and so on until no other event may be added.

Now we show that for K-dense occurrence nets, $CC(N)$
coincides with the collection $L(N)$ of
closed sets of $N$ generated by the $co$ relation. 

Let $N=(B,E,F)$ be a K-dense, interval-finite and degree-finite occurrence net.
\begin{theorem}
Let  $Y \subseteq B \cup E$.
Then $Y \in CC(N)  \iff  Y \in  L(N)$
\end{theorem}
The theorem is an immediate consequence of the following
propositions.
\begin{proposition}\label{p:tagli_Kdensita}
Let $A$ be a B-coset of $N$.
Then $\phi(A)=A^{\ort\ort}$.
\end{proposition}

\begin{proposition}
Let $Y \in CC(N)$, and $A \subseteq Y$ be a B-cut of the subposet induced by $Y$.
Then $\phi(A)=Y$.
\end{proposition}

Therefore, in the case of  K-dense, interval-finite
and degree-finite occurrence nets, the families of
causally closed sets and of closed sets coincide.
In particular, any closed set can be uniquely obtained
starting from a B-coset by applying to it the inductive construction as in
Definition~\ref{d:costruzione_Ai}.

\section{Conclusion}
The main contribution of this paper is twofold.
On one hand, we have applied to locally finite
partially ordered sets a construction which associates
a complete lattice to a given poset;
the elements of the lattice are certain subsets of
the poset, precisely the closed subsets of a closure
operator, defined starting from the \emph{co} relation.
We have shown that, if the partially ordered set satisfies 
a property of local density, i.e.: N-density, 
then the associated lattice is also orthomodular.
Orthomodular posets are studied in the frame of
quantum logic (see, for example, \cite{ql_dcg}).
This suggests to interpret closed sets as propositions in a logical language.

On the other hand, we have focused attention on occurrence
nets as models of concurrent computations, and defined
a family of subsets of the elements of an occurrence net;
we call those subsets \emph{causally closed} because they
can be seen as subprocesses of the whole net which are,
intuitively, closed with respect to the (forward and backward) firing rule
of the net.
We have shown that, when the net is K-dense, the causally
closed sets coincide with the closed sets induced by the
closure operator defined starting from the
\emph{co} relation.

Starting from these first results, we intend to pursue the investigation
of lattices of subprocesses in different directions.
On one hand, we will study further properties of such lattices,
and their relations with domain theory.
On the other hand, we will extend the construction to cyclic
Petri nets, in which a sensible concurrency relation can
be defined even in the absence of a global partial order.

\section*{Acknowledgments}
Work partially supported by MIUR.
\bibliography{bpr_dcm09}
\bibliographystyle{eptcs} 
\end{document}